\documentclass[aps,pre,twocolumn,showpacs,floatfix,preprintnumbers,amsmath,amssymb]{revtex4}
\usepackage{epsfig}
\usepackage{graphicx}
\usepackage{dcolumn}
\usepackage{bm}

\usepackage{amsmath}
\usepackage{amssymb}
\usepackage{tabularx}
\usepackage{mathrsfs}
\usepackage{rotating}

\newcommand{\topcaption}{%
\setlength{\abovecaptionskip}{1pt}%
\setlength{\belowcaptionskip}{0pt}%
\caption}

\begin{document}
\title{Long division unites - long union divides, a model for social network evolution}

\author{J. Jiang$^{1, 2, 3, 4}$}
\email{jiangj2007010209@gmail.com, Telephone:18007155327}
\author{R. Wang$^{2, 5}$}
\author{M. Pezeril$^{3}$}
\author{Q.A. Wang$^{2, 3}$}

\affiliation{\\$^{1}$Complexity Science Center and Institute of
Particle Physics, Hua-Zhong (Central China) Normal University,
Wuhan 430079, China \\$^{2}$Institut Sup\'erieur des Mat\'eriaux
et M\'ecaniques Avanc\'es 44, Avenue F.A. Bartholdi, 72000 Le
Mans, France\\$^{3}$IMMM, UMR 6283, Universit\'e du Maine, Ave. O.
Messaen, 72085 Le Mans, France
\\$^{4}$College of Mathematics and Computer Science of Wuhan
Textile University, Wuhan, 430200, China \\$^{5}$College of
Infromation Science and Engineering, Huajiao University, Quanzhou,
362021, China}

\begin{abstract}
A remarkable phenomenon in the time evolution of many networks
such as cultural, political, national and economic systems, is the
recurrent transition between the states of union and division of
nodes. In this work, we propose a phenomenological modeling,
inspired by the maxim "long union divides and long division
unites", in order to investigate the evolutionary characters of
these networks composed of the entities whose behaviors are
dominated by these two events. The nodes are endowed with
quantities such as identity, ingredient, richness (power),
openness (connections), age, distance, interaction etc. which
determine collectively the evolution in a probabilistic way.
Depending on a tunable parameter, the time evolution of this model
is mainly an alternative domination of union or division state,
with a possible state of final union dominated by one single node.
\end{abstract}


\maketitle

\def\tc{T_{\rm cr}}
This work is a network modeling of the social systems composed of
a large number of entities in interaction whose existence is
dominated by two major events: union and division. Union means
unification of two or more nodes (of the network) into one.
Division is the inverse process: one node splits into several
ones. Union and division is one of the most visible social,
economic, cultural and political phenomena in the course of the
development of a large number of composite systems. Different
countries, political or economic groups can be unified into one
country or group. There are also plenty of examples of division of
these unities. The recurrent character of this phenomenon is well
summarized in the maxim ``long union divides, long division
unites'', consequence of the interplay of a pair of opposite
tendencies in many evolutionary systems.

Union and division are veritable complex processes in which many
factors are responsible of the consequences. A good example of
this complexity is the cultural landscape of the world with the
long history of birth, death and mutation of cultures through
interaction (communication, influence etc.), unification and
splitting of cultures which are in addition under very complicated
and uncertain influence of demographic, genetic, economic,
political, scientific and technological systems as well as of many
accidental elements such as natural disasters, wars, environmental
changes and so on. It is for this reason that the most suitable
description of the stochastic evolution of this kind of systems is
probabilistic modeling taking into account as many as possible the
involved factors and interaction mechanisms.

We present here a phenomenological modeling of networks in which
union and division of nodes are two dominating events determined
in a probabilistic way by the nature and some general features of
the nodes. One of the aims is to see what would be the destiny and
the evolution characters of a network composed of nodes endowed
with some general attributes allowing and influencing unification
and division under given conditions, without enter into the
fundamental principles, interactions and true mechanisms of the
dynamics of the systems. The definition of network in this work
has been inspired by the previous works on cultural networks
\cite{cc}-\cite{CPL1} where a culture is represented by a vector
whose components are quantified by a limited number of values
characterizing cultural features. The vectorial representation of
the nodes certainly applies to many other systems. For example, in
an economic network formed by companies (nodes), each company has
several essential features such as richness or capital, size,
activity domain, age, diversity, openness and so on. In a
political network composed of parties and groups, each political
group has its own characteristic features, the same in the
worldwide networks formed by the countries, so on and so forth.

The phenomenological model of union and division has been used
recently in a dual modeling of political opinion
networks\cite{Jul} in which union and division are two dominating
events happening in the subnetwork of political parties. In
present work, we are interested in knowing the behavior of a
network dominated by the interplay of union/division which occurs
as a function of the natures and features of the nodes. In doing
this, we have in mind some questions about, for example, whether
or not it is possible for all nodes of a given network under
reasonable conditions to merge into one, or what will be the
equilibrium or stationary state in which the number of nodes is
relatively unchangeable, and what are the necessary conditions if
any for these evolutionary behaviors. We think that these
questions are currently of interest because the globalization of
economy, culture and politics becomes a more and more hot topic
and raises different opinions in the world\cite{ghc,rf,CPL2}. We
hope that this modeling of social systems, in spite of its
simplification of the real world where every network is more or
less open to the influence of others, can be helpful for the
understanding of some aspects of the complex world around us and
constitutes a starting point for further work with more realistic
models.

Again, we would like to emphasize that in this probabilistic
model, the mechanism of union and division is not imposed
deterministically to the evolution. At each step of evolution,
union or division may or may not happen, depending on the
probability of occurrence (union, division or nothing) calculated
according to the natures and features of the nodes.

Each node represents one social entity such as country, culture,
political party, company etc. A link between two nodes represents
the interaction between two entities. In the same way one
characterizes, for example, a country or a culture by using the
features such as language, religion, art, custom, political regime
and economy, a quantity named "identity" is introduced to
characterize the nodes. Each node is randomly given an identity
which is represented by a position in a $w$ dimensional space:
$\overset{\rightharpoonup }{c}=\left(c_1,c_2,c_3,\ldots
c_w\right)$($0\leq c_i\leq 1$) with $i=1,2,\ldots w$ where $w$ is
the number of characteristic features we considered. In the
simulation, we set $w=5$. The identity distance between two nodes
$i$ and $j$ is given by:

\begin{equation}
d_{ij}=\sqrt{\text{$\Delta $c}_1{}^2+\text{$\Delta
$c}_2{}^2+\ldots+\text{$\Delta $c}_w{}^2},
\end{equation}
where $\text{$\Delta $c}_1=c_{\text{i1}}-c_{\text{j1}}$. This
distance represents the difference between nodes; similar nodes
have small distance.

A quantity "richness" is introduced to characterize the level and
power of development and/or future cultural, economic and
ideological production and growth. For a culture, this may be the
size and the power of of its content (religion, art, language,
literature, economy, education, sciences and technology etc.), the
population of its carriers and so on. For a country, it can
represent economic, political, cultural, military powers, life
condition as well as the population etc. Larger value of richness
represents higher level of development and power. We suppose that
the richness is globally an increasing function of time. This is
reasonable from the statistical point of view. In the present
model, the time evolution of richness for a given node is given as
follows:
\begin{equation} \label{r}
r=I(a-1),(a\neq1)
\end{equation}
where $I$ is its ingredient and $a$ its age. The ingredient of a
node is introduced to characterize the diversity of a node. It is
defined as the number of previous nodes that have merged into it.

The degree or number of connections of a node is a very important
evolutionary feature. It characterizes the communication state,
the openness, the ability and will of giving/receiving
information, of the node. Nodes with larger degree have more
chance to evolve than smaller degree and isolated nodes. The age
of a node is the time period from its birth to the present time.

The two main processes of the time evolution are merging and
splitting. If two nodes are connected, they can merge into one new
node with probability $p_m$ proportional to the sum of their
degrees, to the difference of their richnesses $\Delta r$ and
inversely proportional to the identity distance $d_{ij}$, that is
\begin{equation}
 p_m (i,j) = A(k_i  + k_j )(\Delta r_{ij}  +
1)/d_{ij},
\end{equation}
where $A$ is the normalization constant, $i$ is the index of the
randomly selected node and $j$ the index of its neighbor. $p_m$ is
normalized over all the neighbors of node $i$. In general, more
the two linked nodes are open, more likely they are close to each
other through the direct and indirect communications (links), and
more likely they merge into one. After merging, a new node $n$ is
born as a composite one characterized by the "ingredient" $I$. The
total ingredient is then equal to the initial size of network $N$
and is conserved. The age of the new composite node is one. Its
richness is supposed to be the sum of the richnesses of the two
merged nodes:
\begin{equation}
r_n=r_i+r_j,(a=1)
\end{equation}
implying a higher level of development and power of the composite
nodes. We take the average value of two merged identities as that
of the composite node. It is computed as follows:
\begin{equation}
\overset{\rightharpoonup
}{c}=\frac{c_{1i}+c_{1j}}{2}\overset{\rightharpoonup
}{1}+\ldots+\frac{c_{\text{wi}}+c_{\text{wj}}}{2}\overset{\rightharpoonup
}{w}.
\end{equation}
The degree of the composite node $n$ is given by
\begin{equation}
k_n=k_i-1+k_j-1-N_{common},
\end{equation}
where $N_{common}$ is the number of common neighbors of node $i$
and $j$. In other words, the evolution of the network after
merging is that all of the neighbors of merged node $i$ and $j$
will connect with the new composite node $n$.

In our model, a randomly selected composite node with ingredient
$I$ can split into $I$ new nodes with a probability $p_s$ given
by:
\begin{equation}    \label{splitprob}
p_s=\frac{1}{Z_s}\frac{Ia(k+k_c)}{r}.
\end{equation}
$p_s$ is normalized by $Z_s=I_{max}a_{max}k_{max}/r_{min}$. We
call this kind of splitting complete splitting. This splitting
probability implies that: 1) composite nodes with larger
ingredient are more likely to split; 2) older nodes are more
likely to split than younger ones; 3) more communication and
openness (larger $k$) facilitates the splitting; 4) nodes with
higher level of development and power (richness) are less likely
to split.

Notice that $k_c$ in Eq.(\ref{splitprob}) is an important tunable
parameter for the dynamics. It implies that even an isolated
composite nodes ($k=0$) can have a non zero chance to split if
$k_c\neq 0$. This kind of single isolated nodes are quite frequent
to the end of a union-dominating period (see Section 3 below) when
the whole network is reduced to only one or two very big (large
ingredient and richness) nodes. If $k_c=0$, the evolution of the
network possibly ends up with a single powerful node. This would
be the definitive disappearance of the network and an eternal
union. This point may be a philosophy-laden subject. Here we
suppose that any single composite node should have a tendency of
splitting driven by the opposite aspects of the different
ingredients already unified in it, or by some intrinsic division
forces. Hence we fixed $k_c=1$ in this work. But a future study of
the evolution with different values of $k_c$ may be interesting
since there may be an interplay between the splitting tendency of
large $k_c$, $I$ and $a$ and the union tendency of large richness
$r$ according to Eq.(\ref{splitprob}).

After splitting, new nodes will inherit the richness from the
split node. We suppose that the richness of each new node is the
same as that of the split node. The identity of the $I$ new nodes
has the same dimensional number and range as the split node, with
randomly given feature values. The age of all new nodes is equal
to one. In addition, the evolution of the network after splitting
is that these new nodes are randomly connected with each other and
with the neighbors of the split node with a constant probability
$p_{rc}=0.9$.

It is worth noticing two conditions of time evolution in our
simulation: 1) at a given time step, a composite node which has
just been formed or a new node which has just split from a
composite node must not be considered as candidate for another
merging or splitting; 2) during the random selection of the nodes,
if the chosen node has $I=1$, the only possible event is merging.
But if the chosen node is a composite one, merging and splitting
are both possible. The decision is random with 0.5 probability for
each possible event.

The initial condition of the simulation is a large number $N=2000$
of nodes randomly connected with a constant probability $p_c=0.3$.
The nodes remaining unconnected will be selected randomly at each
time step and connected to other nodes according to the principle
of preferential attachment \cite{BA}. When $t=1$, the age of each
node is $a=1$, each node has a richness $r=1$ with ingredient
$I=1$.

Next, we will show some numerical results. The time evolution
shows first of all a quasi-periodic behavior, an alternative
domination of union and division state. For example, at the end of
the first cycle, the number of nodes changes from the initial size
$N=2000$ to $N=1$. This is defined as one cycle. Due to the non
zero probability of splitting, the single node at the end of a
cycle will sooner or later split. A new cycle begins. It is found
that the network behaviors are similar in each cycle. The time
evolution of different quantities we studied is described below.

\begin{figure} [ht]
\includegraphics[scale=.4]{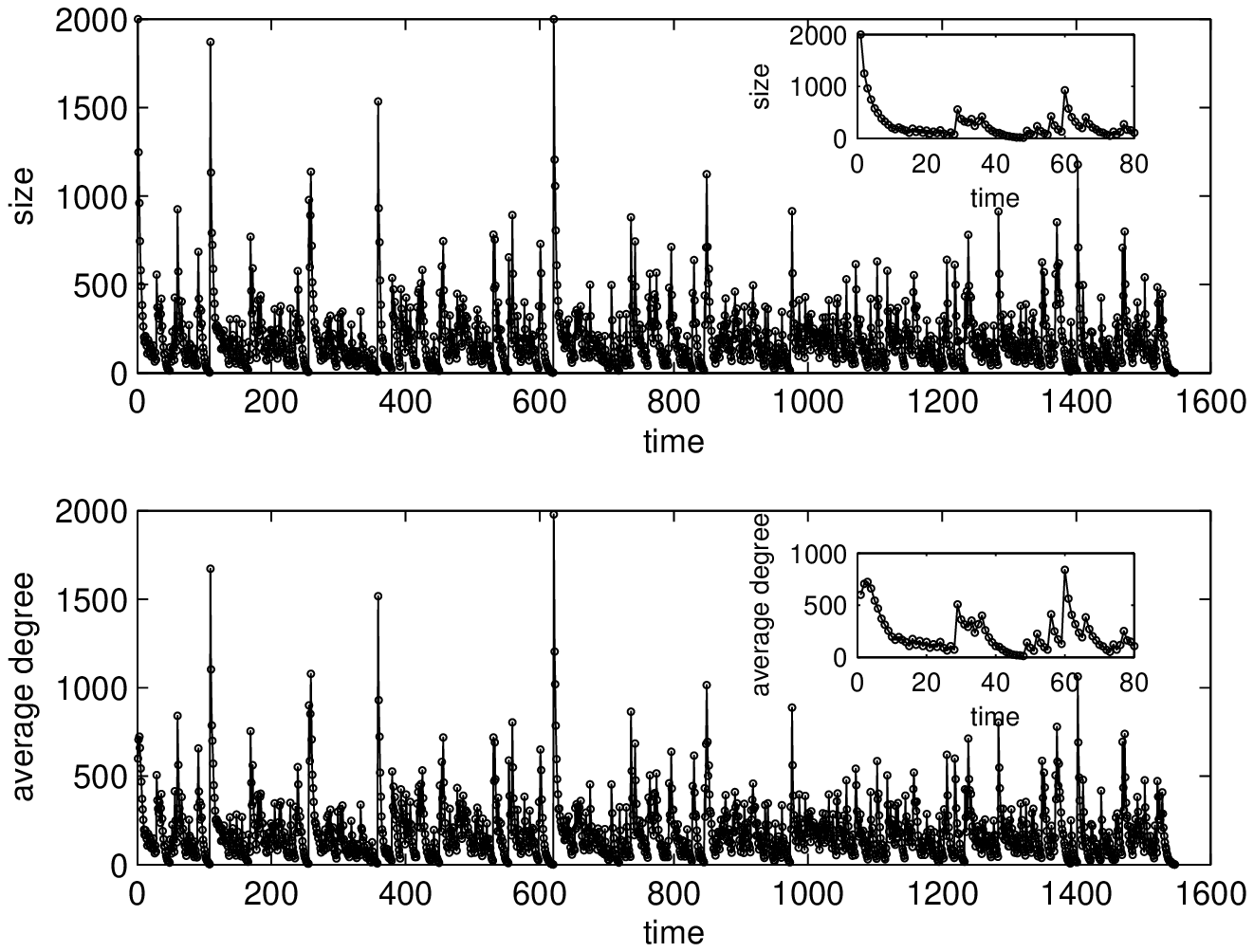}
\centering \topcaption{The time evolution of network size (the
number of nodes in network) (top) and average degree (bottom). The
meaning of X-axis "time" is time steps in the evolution. The size
of network shows large variation. The size at $t=1$ and $621$ has
the maximum value $N=2000$, corresponding to the beginning of the
first and second cycle. When $t=620$ and $1547$, the system size
has the minimum value $N=1$, which corresponds to the end of the
two cycles. In the bottom picture, the time variation of the
average degree shows similar behavior to that of size. The insets
show local amplification of two evolutions.} \label{sizeavedeg}
\end{figure}

The time evolution of the network size (the number of nodes in
network) and the average degree (the average number of each node's
neighbors) are shown in Fig.(\ref{sizeavedeg}). The top picture
shows dramatic variation in the size. We find the largest size
$N=2000$ at $t=1$ and $621$, i.e., at the beginning of the first
and second cycle, respectively. The smallest size $N=1$ occurs for
$t=620$ and $1547$ when these two cycles end. The merging and
splitting processes are responsible for the oscillatory behavior
of size evolution. In the bottom picture of
Fig.(\ref{sizeavedeg}), the average degree evolution is similar to
that of size evolution. This is because the size and the total
degree of the network both increase or decrease during the
division- or union-dominating period. Examples of the periodic
transition between union and division states do not lack in
historical facts: the union and division of nations, dynasties,
and economic or financial groups, etc.

\begin{figure} [ht]
\includegraphics[scale=.4]{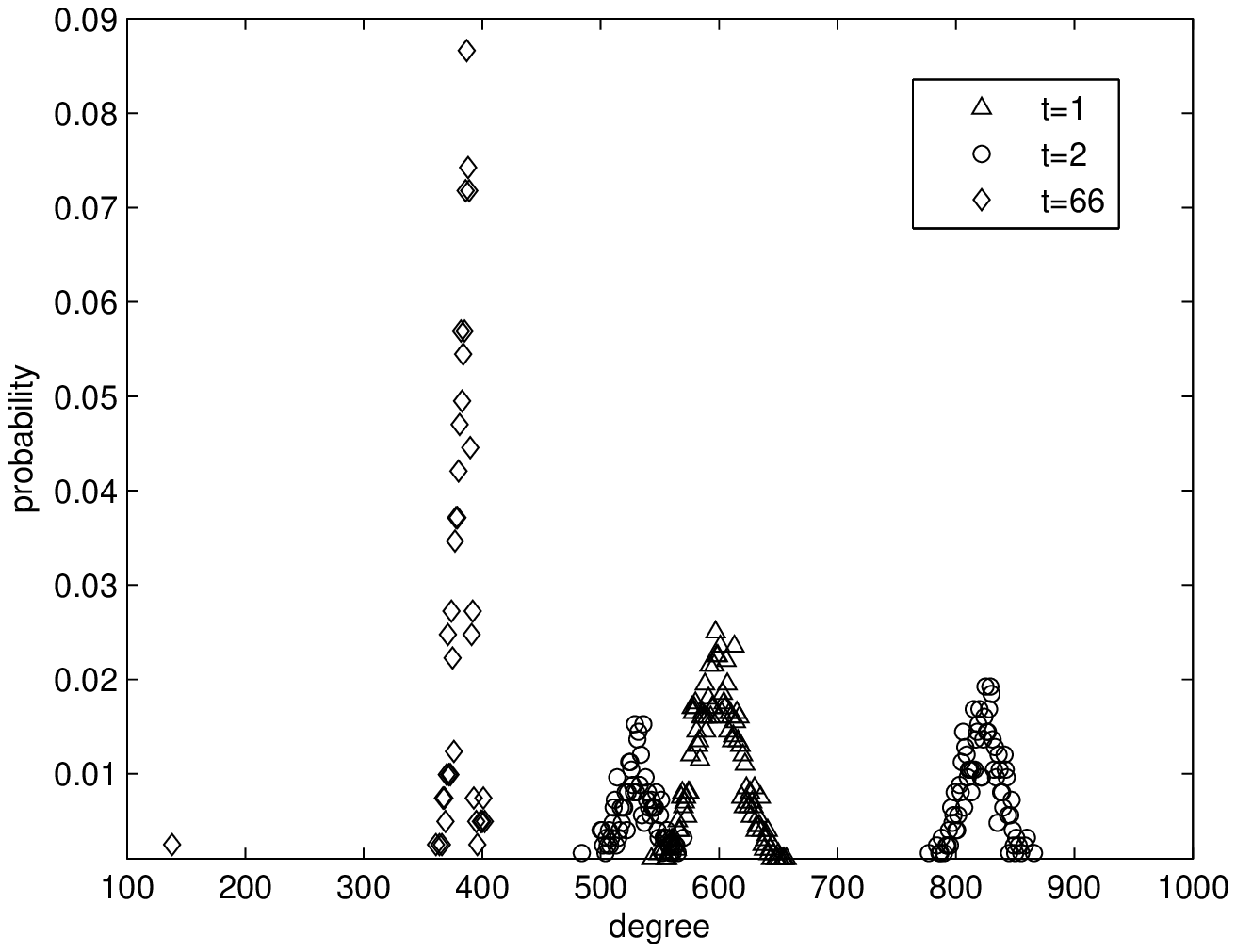}
\centering \topcaption{Degree distributions at different time
steps, $t=1$ (triangle), $t=2$ (circle), and $t=66$ (diamond) in
the first cycle. The distribution at $t=1$ is a Poisson
distribution with a sharp peak since the initial network is random
graph. $t=2$, the sharp distribution is separated into two
discontinuous distributions with larger or smaller degree. The two
peaks merge into one at $t=66$, yielding another sharp
distribution.} \label{degdis}
\end{figure}

In Fig.(\ref{degdis}), the degree distributions are given at $t=1$
(triangle), $t=2$ (circle) and $t=66$ (diamond). We see that the
initial distribution ($t=1$) is a Poisson distribution with very
sharp peak due to the random connections in the initial situation.
At $t=2$, after merging (no splitting happens at $t=1$ as no
composite node exists), the distribution of one peak splits into
two peaks: one is composed of the composite nodes with larger
degree (the number of connection is increased from the merged
nodes' neighbors), another peak with smaller degree is composed of
the composite nodes and the neighbors of merged nodes. The
discontinuity of these two distributions means that the nodes are
separated into two distinguished classes. In the later evolution,
the merging in the group with big degree nodes causes the shift of
the distribution to the smaller degree region. The shift is faster
for the larger degree peak than for the smaller degree peak. The
reason is that, according to the merging probability, the nodes
with larger degree have larger merging probability than those with
smaller degree all things being equal else. As a consequence, the
two peaks merge into one at $t=66$ with a very sharp distribution.
Hence during the whole evolution, when the dominated process is
merging, the degree distribution undergoes a shift from large
degree to small degree region. Inversely, when the dominated
process is splitting, the shift is in opposite direction.

\begin{figure} [ht]
\includegraphics[scale=.4]{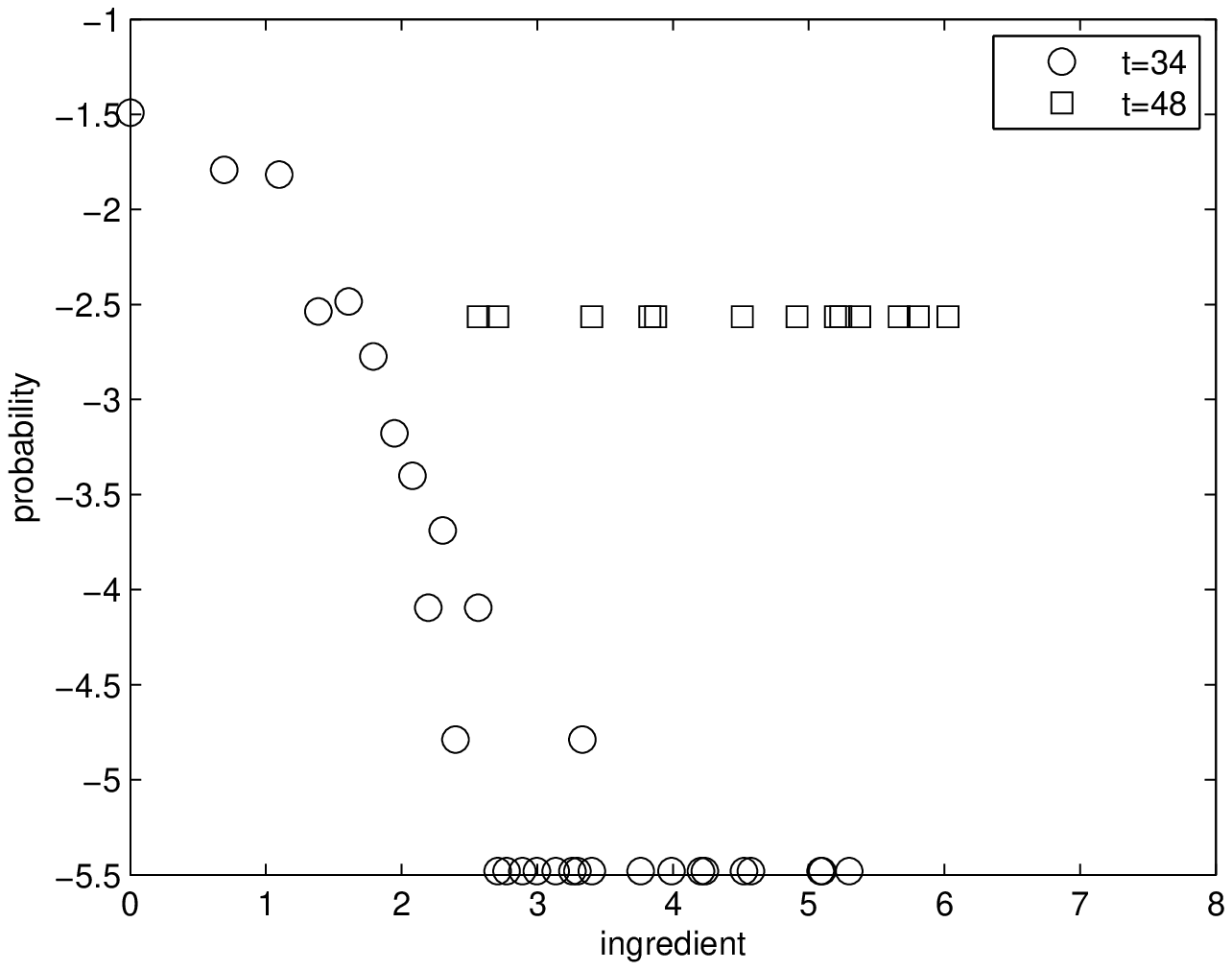}
\centering \topcaption{The ingredient distribution at $t=34$
(circle) and $t=48$ (square) with double-log scaling. The former
distribution is close to power law with fat tail and the latter is
a uniform one, which are the major distributions clearly observed
during the evolution. When the merging is dominating, the
distribution changes from power law to uniform one. When the
splitting is dominating, the transition is inverted.}
\label{ingdis}
\end{figure}

The ingredient distribution is shown in Fig.(\ref{ingdis}) with
double-log scaling. Due to the merging, the ingredient of the
composite node becomes larger than one and increases with time.
There is no obvious regularity in its distribution at the first
stages of the evolution, since the ingredient values are small. At
$t=34$, the distribution is close to a power law with fat tail
marked by circle points, which implies that most of nodes have
small ingredient, and few nodes have very large ingredient. Before
$t=48$, the merging is the dominating process resulting in the
decreasing of nodes with small ingredient and the increasing of
nodes with large ingredient. When $t=48$, the power law
distribution evolves into a uniform one marked by square points,
different ingredients have the same probability to appear. In the
following period, the splitting of the nodes with large ingredient
is dominating and results in the increasing of nodes with small
ingredient. The uniform distribution gradually evolves into power
law distribution again. This transition of ingredient distribution
between the power law and the uniform one is repeated
continuously.

\begin{figure} [ht]
\includegraphics[scale=1.2]{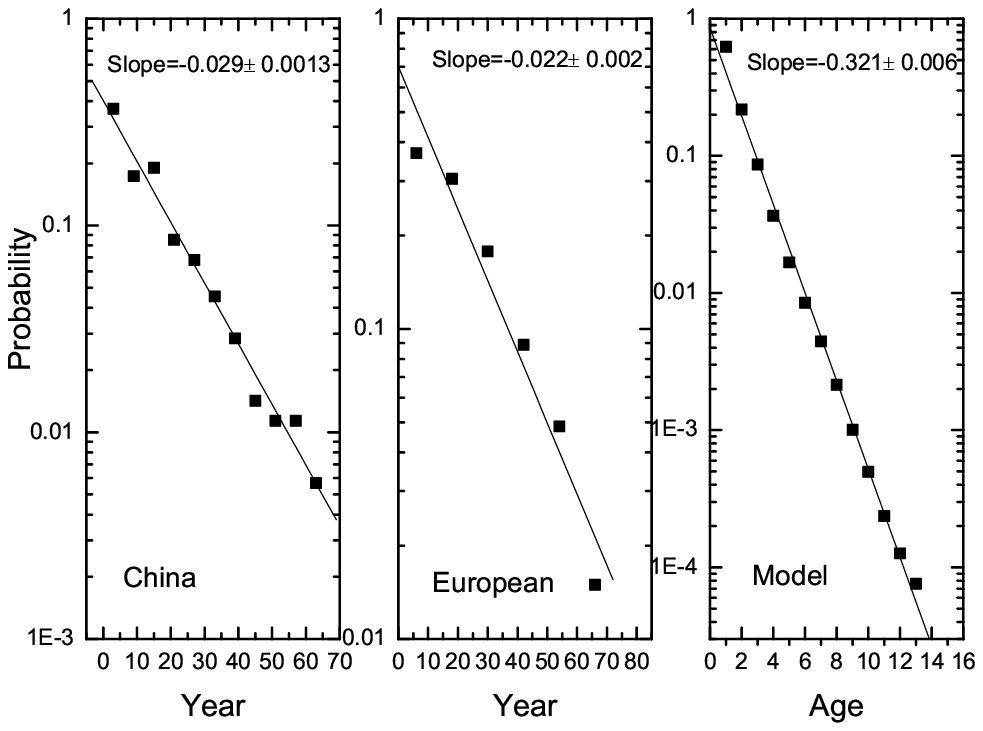}
\centering \topcaption{Comparison of the lifetime distribution of
our model with the empirical data from different kingdoms and
dynasties in Europe and China. Three distributions are all
exponential (straight line in log-linear plot). The slopes of each
line are: $-0.029\pm 0.0013$ $year^{-1}$(China), $-0.022\pm 0.002$
$year^{-1}$ (Europe), and $-0.321\pm 0.006$ $step^{-1}$ (our
model).} \label{age}
\end{figure}

In Fig.(\ref{age}) the lifetime (duration of existence of the
nodes) distribution of this model is compared with two empirical
ones we have established from the data of different kingdoms and
dynasties in China \cite{china} and in Europe \cite{european}. The
distributions are exponential (straight line in log-linear plot).
The slopes of the straight lines are: $-0.029\pm 0.0013$
$year^{-1}$(China), $-0.022\pm 0.002$ $year^{-1}$ (European), and
$-0.321\pm 0.006$ $step^{-1}$ (our model). The agreement between
the numerical result and the real data allows to determine the
time scale of the model when it is applied to this kind of
systems. For example, we find that the average lifetime of the
model, the dynasties in China and in Europe are 2.1 steps, 14.3
years, and 19.4 years, respectively. This suggests that one time
step in our model is equivalent to 6.8 years in the evolution of
the dynasties of China, and 9.2 years in that of the European
dynasties.

In summary, we proposed a phenomenological model of evolution of
social networks with the aim of investigating the behavior and the
fate of the networks whose nodes evolve with these two dominating
events. This work revealed interesting evolutionary features, some
of them being not expected before the simulation. For example, we
expected stationary or equilibrium state of evolution with more
than one nodes and constant values of certain quantities such as
network size, degree etc. But this state seems impossible in
present model. The evolution either terminates in a final union
with the disappearance of the network and a single node, or never
ceases a nonequilibrium and periodic evolution with fluctuating
oscillation of network size and other quantities. The main
features of the model can be summarized as follows:

1) The degree distribution is Poisson distribution at the
beginning of the evolution. After this, there is a transition
between a single continuous distribution and a discontinuous one
with two peaks which means a separation of the nodes into two
classes of connection state (openness) with large and small
degree, respectively. 2) The ingredient distribution undergoes a
periodic transition between power law distribution and uniform
one. 3) The lifetime distribution is in exponential form. A
mapping of time scale from our modeling to real systems could be
made for the kingdoms and dynasties in China and Europe by
comparing the numerical distribution to empirical data of lifetime
we collected.

Besides the mechanisms of evolution reported in this
paper, we also tried many others in order to compare their
results. For example, we considered the age of links defined as
the elapsed time from the born of a connection to its
disappearance. We also considered simpler probability of merging
proportional to the sum or the difference of two nodes' degrees.
The probability of splitting was also simplified to a simple
proportion to the age of node. The results of the simulation with
these different mechanisms are close to what we reported here with
similar periodic fluctuating evolution of the network.

At last, we want to emphasis that the applicability of this model
is limited with the social network where the local interaction is
dominated in the evolution. It is not suitable for networks
companied with global interaction or noise (i.e. random
influences, which lead to the feature of nodes or dynamical
evolution changes with a small probability). So further
improvement of the model is possible in these networks. For
instance, the merging of any pair of nodes or the splitting of any
node may take place randomly with a small probability by some
external influences.


\begin{thebibliography}{99}
\bibitem {cc}
Castellano C, Fortunato S and Loreto V, 2009 \emph{Rev. Mod.
Phys}. \textbf{81} 591 \%DOI:10.1103/RevModPhys.81.591

\bibitem{RA}
Axelrod R, 1997 \emph{J. Conflict Res}, \textbf{41} 203
\%DOI:10.1177/0022002797041002001

\bibitem{MNK}
Kuperman M N, 2006 \emph{Phys. Rev. E}, \textbf{73} 046139

\bibitem{NL}
Lanchier N, 2010 arXiv:1004.0365v1

\bibitem{KK}
Klemm K, Eguiluz V M, Toral R and Miguel M S, 2003 \emph{Physica
A}, \textbf{327} 1 \%DOI:10.1016/S0378-4371(03)00428-X

\bibitem{KK2}
Klemm K, Eguiluz V M, Toral R and Miguel M S, 2003 \emph{Phys.
Rev. E}, \textbf{67} 026120

\bibitem{KK3}
Klemm K, Eguiluz V M, Toral R and Miguel M S, 2003 \emph{Phys.
Rev. E}, \textbf{67} 045101

\bibitem{KK4}
Klemm K, Eguiluz V M, Toral R and Miguel M S, 2005 \emph{J.
Economic Dynamics Contral}, \textbf{29} 321
\%DOI:10.1016/j.jedc.2003.08.005

\bibitem{DC}
Centola D, Avella J C G, Eguiluz V M and Miguel M S, 2007 \emph{J.
Conflict Res}, \textbf{51} 905 \%DOI:10.1177/0022002707307632

\bibitem{RT}
Toral R and Tessone C J, 2007 \emph{Communications in
Computational Physics}, \textbf{2} 177
\%DOI:http://www.arxiv.org/abs/physics/0607252

\bibitem{TSE}
Evans T S and Plato A D K, 2007 \emph{Phys. Rev. E}, \textbf{75}
056101

\bibitem{JCGA}
Avella J C G, Eguiluz V M, Miguel M S, Cosenza M G and Klemm K,
2007 \emph{Journal of artificial societies and social simulation},
\textbf{10} 9 \%DOI:http://jasss.soc.surrey.ac.uk/10/3/9.html

\bibitem{JCGA2}
Avella J C G, Eguiluz V M, Cosenza M G, Klemm K, Herrera J L and
Miguel M S, 2006 \emph{Phys. Rev. E} \textbf{73} 046119

\bibitem{cc2}
Castellano C, Marsili M and Vespignani A , 2000 \emph{Phys. Rev.
Lett} \textbf{85}, 3536

\bibitem{CGL}
Lazaro C G, Lafuerza L F, Floria L M and Moreno Y, 2009
\emph{Phys. Rev. E} \textbf{80} 046123

\bibitem{DV}
Vilone D, Vespignani A and Castellano C, 2002 \emph{Eur. Phys. J.
B} \textbf{30} 399

\bibitem{CPL1}
Sun K, Mao X M and Ou Y Q, 2002 \emph{Chinese Physics B},
\textbf{11} 1280

\bibitem {Jul}
Wang R and Wang Q A, A dual model for opinion networks, 2011
\emph{Phys. Rev. E}, \textbf{84} 006100

\bibitem{ghc}
Cornwell G H and Stoddard E W, 2001 \emph{Global Multiculturalism:
Comparative Perspectives on Ethnicity, Race, and Nation}, Rowman
\& Littlefield Publishers

\bibitem{rf}
Falk R, 1999 \emph{Predatory Globalization: A Critique}, Princeton
University, Polity Press \%DOI:10.1177/146499340100100112

\bibitem{CPL2}
Chang Y F and Cai X, 2007 \emph{Chinese Physics Letters},
\textbf{24} 2430

\bibitem{BA}
Barabasi A L and Albert R, 1999 \emph{Science}, \textbf{286} 509


\bibitem{china}
http://baike.baidu.com/view/224457.htm

\bibitem{european}
http://zhidao.baidu.com/question/10662233.html

\end{thebibliography}
\end{document}